# Best-in-class and Strategic Benchmarking of Scientific Subject Categories of Web of Science in 2010

J. A. García, Rosa Rodriguez-Sánchez, J. Fdez-Valdivia, Nicolas Robinson-García and Daniel Torres-Salinas

**Abstract** Here we show a novel technique for comparing subject categories, where the prestige of academic journals in each category is represented statistically by an impact-factor histogram. For each subject category we compute the probability of occurrence of scholarly journals with impact factor in different intervals. Here impact factor is measured with Thomson Reuters Impact Factor, Eigenfactor Score, and Immediacy Index. Assuming the probabilities associated with a pair of subject categories our objective is to measure the degree of dissimilarity between them. To do so, we use an axiomatic characterization for predicting dissimilarity between subject categories. The scientific subject categories of Web of Science in 2010 were used to test the proposed approach for benchmarking Cell Biology and Computer Science Information Systems with the rest as two case studies. The former is best-in-class benchmarking that involves studying the leading competitor category; the latter is strategic benchmarking that involves observing how other scientific subject categories compete.

# Introduction

During the last decades the evaluation of research activity by means of bibliometric methodologies has widely expanded. Researchers are more than ever immersed in a demanding "publish or perish" culture, fueled not just by researchers' competitiveness, but also by national research assessment agencies which have become more and more frequent (Jiménez-Contreras et al., 2003; Abramo et al., 2011) as a means to favor countries and universities' improvement in terms of research performance.

Although these exercises are conceived for enhancing research excellence, they seemed to prompt quantity rather than quality at a first stage (Moed, 2008). However, over the last decade a shift in researchers' publication behavior can be observed. In this sense, many studies suggest a greater demand on publishing in high-ranked journals as the means to reaching academic excellence and success rather than just focusing on quantity (Leahey, 2007). But researchers' efforts to publish in reputed journals are not only the consequence of the introduction of bibliometric indicators, but also their historical need for acknowledgment and prestige through their works (Luukkonen, 1992).

Although some malpractices have been found due to this obsessive need to publish in highly ranked journals (Fanelli, 2010), researchers ambition to gain the greatest visibility and hence, a greater chance to have more impact, seems to be completely legitimate. In this sense, the Journal Impact Factor (hereafter IF) has played a key role as judge, not only of journals' prestige through Thomson Reuters' Journal Citation Reports (hereafter JCR), but also of national research assessment exercises focused on distributing research funding (Jiménez-Contreras et al., 2003; Adam, 2002). Although IF was not originally intended for this use, it is considered a proxy for research competitiveness along with other indicators related to research impact and visibility, and it has become the main criterion when ranking journals. However, many critical voices have emphasized many of its shortcomings and limitations when using IF for such purposes.

In this line of thought, Leydesdorff & Bornmann (2011) discuss that the IF may be misleading when measuring impact as citation curves are highly skewed. Moed (2008) also points out a well discussed limitation related to the different citation patterns each research field has which would benefit journals from certain areas in which the citation rate is higher from those in which the citation rate shows lower figures. In this sense, the JCR tries to solve this latter limitation by dividing journal rankings according to subject categories which are used as proxy for research fields.

However, such approach also shows some shortcomings which may be taken into account. Because these journal rankings are usually divided into quartiles (Garcia et al, 2012a) considering as highly ranked those journals belonging to the first quartile, it is easy to assume that different journals positioned in the first quartile for different subject categories should have similar impact. But this assumption may be questionable as, on the first hand, the division between subject categories is not always clear, in fact it is common to find many journals categorized in more than one, and secondly, research is becoming more interdisciplinary in certain areas (Buter et al., 2011). Therefore, in some cases, researchers may be interested on publishing in journals belonging to a similar subject category to that which would encompass their line of work, but which may contain journals with a higher impact, consequently gaining more visibility.

In line with a previous study in which subject categories where ranked according to their multidimensional prestige (Garcia et al, 2012b), in this paper our target is to apply a novel methodology for benchmarking subject categories. This methodology is based on information gain or Kullback-Leibler divergence (Kullback & Leibler, 1951), in which distributions of a given indicator are compared meaning that the more similar they are, the lesser information gain there is between them. This methodology has already been successfully tested for comparing academic institutions (García et al., 2012c) and citation patterns of book chapters (Torres-Salinas et al., 2013). In this approach, the prestige of academic journals in each category is represented statistically by an impact-factor histogram, and thus, we compute the probability of occurrence of scholarly journals with impact factor in different intervals. Here impact factor is measured with Thomson Reuters Impact Factor, Eigenfactor Score, and Immediacy Index.

Specifically we aim at studying if the information gain measure could be a valid indicator to compare categories according to their impact distributions and which impact indicators should be used. Assuming the probabilities associated with a pair of subject categories our objective is to measure the degree of dissimilarity between them using a formal tool. We address the following research questions (RQ):

RQ1 – Can the information gain methodology be applied to compare subject categories in terms of impact similarity?

RQ2 – How do impact indicators reflect the similarity between subject categories?

RQ3 – Is this methodology affected by the interdisciplinarity of subject categories?

The paper is structured as follows. In section 2 we present the data source and the basic methodology for benchmarking scientific subject categories. Next, in section 3 we test our methodology through two case studies. For this, we used the 174 scientific subject categories of Web of Science in 2010 and we selected as case studies Cell Biology as a category highly focused on an enclosed field, and Computer Science Information Systems as a more interdisciplinary field. These two fields were selected as the offer two viewpoints from which one can compare categories. The former case study is best-in-class benchmarking, involving the study of a leading competitor category; the latter is strategic benchmarking, which involves observing how other scientific subject categories compete with Computer Science Information Systems. In section 4 we end with some concluding remarks. Finally, in the Appendix we refer the reader to the axiomatic characterization for predicting dissimilarity between subject categories proposed.

## Data and methods

In order to analyze the degree of dissimilarity between scientific subject categories based on the prestige of their respective journals, the Thomson-Reuters Web of Science database was selected as data source. This decision is based on the great regard this database has for research policy makers, as it is considered to store the most relevant scientific literature in the world. This database publishes every year the Journal Citation Reports (henceforth JCR), which are lists of journals classified by subject categories and which provide 8 different bibliometric indicators (Total cites, IF, 5-year IF, Immediacy Index, Articles, Cited half-life, Eigenfactor Score and Article Influence Score). Next, we explain in detail the retrieval process and briefly describe the motives behind the selection of the two study cases (Cell Biology and Computer Science Information Systems). Then, we give the key points for interpreting the methodology employed, which will be represented by heliocentric clockwise maps. For further information on such methodology, the reader is referred to the Appendix.

**Data source**

We retrieved the data from the 2010 edition of the JCR which is structured into 174 scientific subject categories. For this we downloaded manually the IF, Eigenfactor Scores and Immediacy Index of all journals. The most relevant impact indicator is the JIF, which is often used to rank journals. This indicator is commonly used to measure journals' impact and would be the natural choice in order to estimate the impact-factor histograms over which we calculate the degree of dissimilarity. However, we did not limit our study to this indicator and we also selected the Eigenfactor Scores and the Immediacy Index in order to study differences among each other. Once data was processed, we selected two scientific subject categories (Cell Biology, and Computer Science Information Systems) in order to estimate their information gain values when comparing with the rest of the subject categories.

Garcia et al. (2012b) observed that Cell Biology is the top subject category of a ranking of the 174 scientific subject categories of Web of Science in 2010, based on the measurement of multidimensional prestige of influential journals. The multidimensional prestige of influential journals takes into account the fact that several prestige indicators should be used for a distinct analysis of the impact of scholarly journals in a subject category. After having identified the multidimensionally influential journals, their prestige scores can be aggregated to produce a summary measure of multidimensional prestige for a subject category.

In Garcia et al. (2012b) unsupervised statistical classification was used in order to identify groups of scientific subject categories from the Web of Science in 2010, sharing similar characteristics in the corresponding values of multidimensional prestige. That is, subject categories having the highest, medium, and lowest multidimensional prestige of influential journals. In Table 4 of Garcia et al. (2012b) we observe that Computer Science Information Systems belongs to the class of scientific subject categories having medium multidimensional prestige of influential journals. In conclusion, the selected subject categories (Cell Biology, and Computer Science Information Systems) are very different between each other, making them interesting cases to analyze.

## Methodology for benchmarking scientific subject categories

In this study we benchmark two subject categories from the JCR in 2010. Here, the prestige of academic journals in different subject categories is characterized statistically by impact-factor histograms. Here we will refer as impact factor to the different prestige indicators analyzed, that is, IF, Eigenfactor Score (ES) and Immediacy Index (II).

For instance, we can compute the probability of occurrence of journals with IF values in different intervals. Figure 1 shows the corresponding histograms of IF values which were computed to both Cell Biology and Computer Science Information Systems.

Fig. 2 illustrates the Eigenfactor-Score histograms for representing Cell Biology and Computer Science Information Systems in 2010. In this case we calculate the probability of occurrence of academic journals having different ES values, for each category under analysis.

And Fig. 3 shows the corresponding Immediacy-Index histograms based on the probability of occurrence of academic journals with II values in different intervals, for both Cell Biology and Computer Science Information Systems in 2010.

**Fig. 1** Thomson Reuters Impact-Factor histograms representing Cell Biology and Computer Science Information Systems in 2010

Let us assume the discrete probabilities (histogram) associated with a reference subject category $R$ (e.g., Computer Science Information Systems) and another category of input $I$ as those given by $P$ and $Q$, respectively. In the Appendix we present a basic axiomatic characterization for a measure of information gain between an input subject category $I$ and another of reference $R$, where the information gain measures the degree of dissimilarity between these two subject categories $R$ and $I$, based on the respective histograms $P$ and $Q$.

**Fig. 2** Eigenfactor-Score histograms representing Cell Biology and Computer Science Information Systems in 2010

In this paper, following the results presented in the Appendix, the degree of dissimilarity between two discrete probability distributions $P$ and $Q$ is to be measured using the Kullback-Leibler information function (Kullback, 1978).

Information gain or Kullback-Leibler information function is a measure that allows us to select the subject categories more alike to a given category of reference. It compares two distributions; a true probability distribution $P(x)$ and an arbitrary probability distribution $Q(x)$, and indicates the difference between the probability of $X$ if $Q(x)$ is followed, and the probability of $X$ if $P(x)$ is followed.

**Fig. 3** Immediacy-Index histograms representing Cell Biology and Computer Science Information Systems in 2010

If we predict the dissimilarity between two subject categories (a given reference category $R$ and another category of input $I$) based on their information gain, then the minimum value of information gain between $R$ and any other category of input $I$ leads to the most alike category $I$ to the journal impact distribution of the reference category $R$.

In order to illustrate the information gain values, we have developed what we have called the 'Heliocentric Clockwise Maps' (see Fig. 4). These maps are interpreted as follows. The center of the circle would be the subject category to which the other categories are compared; in our case it would represent the reference subject category (e.g., Cell Biology). The dots surrounding the centre of the circle would represent the input subject categories. Therefore, the ones closer to the center (lower information gain values) would show a more similar journal impact profile (impact factor histogram) to that of the reference subject category and the ones further way (higher information gain values) would perform more differently. The maps are named clockwise because the order of the subject categories represents their multidimensional prestige values starting on the top of the map (see Table 4 in Garcia et al. (2012b)). Therefore, the subject category at the top of the circle has the highest multidimensional prestige and so on, until the one on its left side which shows the lowest multidimensional prestige. This allows the reader to better interpret the meaning of more or lesser information gain (higher multidimensional prestige or lower multidimensional prestige) and the relation between information gain and multidimensional prestige. Only 30 subject categories were considered in the construction of the heliocentric clockwise maps. These are the most similar ones to the one used as case study.

**Fig. 4** Methodology for benchmarking scientific subject categories

# Study Case: Benchmarking Scientific Subject Categories of Web of Science in 2010

In this section we designed an ad hoc heliocentric map, allowing the reader to easily analyze the similarity of the study case scientific subject categories with the rest of scientific subject categories of Web of Science.

We use the Information Gain (i.e., Kullback-Leibler information function) based on three different types of histograms that characterize probabilistically the subject categories according to their impact. In this case, we will compare them in the two scenarios above mentioned with the measurement of multidimensional prestige of influential journals in each subject category proposed by (Garcia et al, 2012b) and where the prestige scores of multidimensionally influential journals are aggregated to produce a summary measure of multidimensional prestige for each category.

**Fig. 5** Heliocentric map representing the Information Gain for Cell Biology in 2010. Subject categories are characterized by Thomson Reuters Impact-Factor histograms.

In Figures 5-10 we show the results obtained for our case studies. In these figures, the reference subject category is positioned in the middle of the heliocentric map (either, Cell Biology or Computer Science Information Systems) and other 30 scientific subject categories are placed around it considering their similarity. These 30 categories are the ones with the lowest values of information gain with respect to the reference subject category. That is, they are the 30 categories most alike to the reference one in each case study. Scientific subject categories are ordered clockwise in the heliocentric map according their multidimensional prestige values starting on the top of the figure (see Table 4 in Garcia et al., 2012b).

**Fig. 6** Heliocentric map representing the Information Gain for Cell Biology in 2010. Subject categories are characterized by Eigenfactor Score histograms.

In Figs. 5 and 8, scientific subject categories of Web of Science are characterized IF histograms; whereas in Figs. 6 and 9, subject categories are represented by ES histograms. Finally, to produce the results illustrated in Figs. 7 and 10, categories of Web of Science were represented using II histograms.

In Figures 5-7, we observe that the most similar scientific subject categories to Cell Biology are life or medical sciences categories with just a few exceptions (e.g., Nanoscience and Nanotechnology; Chemistry Physical). This is best-in-class benchmarking that involves studying the leading competitor, i.e., Cell Biology. It identifies scientific subject categories that are leaders in the JCR for the 2010 edition, using a specific statistical characterization (i.e., the IF, ES, or II histograms).

**Fig. 7** Heliocentric map representing the Information Gain for Cell Biology in 2010. Subject categories are characterized by Immediacy Index histograms.

In these figures, we observe that the configuration has the form of a spiral, where the most alike categories are often in the class with the highest multidimensional prestige of influential journals (see Table 4 in Garcia et al (2012b)) and as we go down in the ranking based on the measurement of multidimensional prestige, subject categories are more dissimilar.

The most similar scientific subject categories to Cell Biology (Biochemistry & Molecular Biology, Neuroscience, Endocrinology & Metabolism, Immunology, Genetics & Heredity, Oncology, Biophpysics, Microbiology, Hematology, Cardiac & Cardiovascular Systems, Biochemical Research Methods) have several characteristics in common; they are life or medical sciences categories, and they are in the class of highest multidimensional prestige. They therefore perform very similarly not just considering their multidimensional prestige of influential journals but also their journal impact profile (histograms), as it is shown by their respective information gain values which are illustrated in the Heliocentric maps.

On the other side, we find that those ranked in the lowest positions of the ranking based on the measurement of multidimensional prestige are the ones with higher information gain, and consequently, more dissimilar with respect to the journal impact profile of the reference subject category.

In Figures 8-10 we show the case of Computer Science Information Systems. Recall that Information Systems belongs to the class of scientific subject categories having medium multidimensional prestige of influential journals (see Table 4 in Garcia et al (2012b)). Therefore this is strategic benchmarking that involves observing how other scientific subject categories compete using a specific statistical characterization.

In this case, due to the medium impact of its journals, there are many categories similar to the journal impact profile (histogram) of Computer Science Information Systems. Considering their information gain values, several Computer Science categories have very similar journal impact profile to Computer Science Information Systems. These are Artificial Intelligence, Interdisciplinary Applications, Theory & Methods, and Software Engineering. However, we also identify other categories that have similar journal impact profile to Computer Science Information Systems but are thematically different (i.e., Statistics & Probability, Zoology, Food Science & Technology, or Mathematics Interdisciplinary Applications).

From Figs. 8-10 we have that Computer Science, Management, Health/Medical Sciences, and Engineering, are similar to the Information Systems' journal impact profile. These results concur with those from Cronin & Meho (2008) which show that Information Studies has become a much more successful exporter of ideas than in the recent past, and it is less introverted than before, drawing more heavily on the literature of such disciplines as Computer Science and Engineering on the one hand and Management on the other.

**Fig. 8** Heliocentric map representing the Information Gain for Computer Science Information Systems in 2010. Subject categories are characterized by Thomson Reuters Impact Factor histograms.

**Fig. 9** Heliocentric map representing the Information Gain for Computer Science Information Systems in 2010. Subject categories are characterized by Eigenfactor Score histograms.

**Fig. 10** Heliocentric map representing the Information Gain for Computer Science Information Systems in 2010. Subject categories are characterized by Immediacy Index histograms.

## Conclusions

In this paper we present a theoretic information measure for benchmarking subject categories. We analyze its usefulness by applying it to the impact factor histograms in two case studies in which we compared a given reference subject category with the rest of scientific subject categories from the 2010 edition of the JCR. Three different types of histograms were used, according to the following indicators IF, ES and II.

The chosen subject categories of reference were Cell Biology and Computer Science Information Systems which are very different between each other according to the multidimensional prestige, making them interesting cases to analyze. Cell Biology is in the top of a ranking of the 174 scientific subject categories of Web of Science in 2010. Computer Science Information Systems belongs to the class of scientific subject categories having medium multidimensional prestige of influential journals. Information Gain is a measure of dissimilarity between discrete probability distributions. It satisfies a number of properties for comparing two subject categories by means of the difference between their impact-factor histograms.

The Information Gain closely relates to similarity between subject categories as perceived by using a different model: The multidimensional prestige of influential journals in each subject category. In conclusion, both theoretical and empirical results imply that it can be used to benchmark subject categories using an information theoretic approach. However, it performs differently according to the level of interdisciplinarity of the subject category selected. In the best-in-class benchmarking of Cell Biology we have identified the scientific subject categories that were leaders in the 2010 edition of JCR: Biochemistry & Molecular Biology, Neurosciences, Endocrinology & Metabolism, Immunology, Genetics & Heredity, Oncology, Biophysics, Microbiology, and Cardiac & Cardiovascular Systems. As can be seen from benchmarking heliocentric maps illustrated in Figures 5-7, it follows that similar results are obtained using any of the three different statistical characterizations for a subject category: IF, ES and II histograms. The most alike categories are often in the class with the highest multidimensional prestige of influential journals in a subject category. Also they are all thematically related, showing a disciplinary coherence when comparing.

In the strategic benchmarking of Computer Science Information Systems we identified several Computer Science categories having very similar journal impact profile to Computer Science Information Systems (in 2010) like Artificial Intelligence, Interdisciplinary Applications, Theory & Methods, and Software Engineering. But other categories have also a similar journal impact profile like Statistics & Probability, Zoology, Food Science & Technology, and Mathematics Interdisciplinary Applications. In this study case (Figures 8-10), we conclude that the use of distinct statistical characterizations (IF, ES, or II histograms) leads to similar results. However, as a more interdisciplinary subject category, there is no a disciplinary coherence within similar subject categories.

**Acknowledgments.** This research was sponsored by the Spanish Board for Science and Technology (MICINN) under grant TIN2010-15157 co-financed with European FEDER funds. Nicolás Robinson-García is currently supported by a FPU grant from the Ministerio de Educaci_on y Ciencia of the Spanish Government. Thanks are due to the reviewers for their constructive suggestions.

Moed, H.F. (2008). UK research assessment exercises: informed judgments on research quality or quantity? Scientometrics, 74(1), 153-161.

Rao, C.R. (1982). Diversity and dissimilarity coefficients: A unified approach. Theoretic Population Biology, Vol. 21, No. 1, 24-43.

Renyi, A. (1961). On measures of entropy and information. Proc. Fourth Berkeley Sump. Math. Stat. and Prob., University of California Press, Vol. 1, 547-561.

Shannon, C.E. (1948). A mathematical theory of communication. Bell System Tech. J., Vol. 27, 379-423; 623-656.

Sharma, B.D., & Mittal, D.P. (1977). New non-additive measures of relative information. Journ. Comb. Inf. Syst. Sci., Vol. 2, 122-132.

Torres-Salinas, D., Rodríguez-Sánchez, R., Robinson-García, N., Fdez-Valdivia, J., & García, J.A. (2013). Mapping citation patterns of book chapters in the Book Citation Index. Journal of Informetrics. In press.

Wiener, N. (1950). The Human use of human beings, Houghton Mifflin Co., Boston.

# Apppendix A: Axiomatic characterization of a measure of information gain between categories

Here we characterize the relative information between subject categories with a minimal number of properties which are natural and thus desirable. Next, it will be derived the form of all information functions satisfying these properties which we have stated to be desirable for predicting differences between subject categories. Thus, a first postulate states a property of how unexpected a single event of a subject category was.

**Axiom 1.** *A measure $U$ of how unexpected the single event "a category's journal with impact factor in the interval $[l, l + \Delta l)$ occurs" was, depends only upon its probability $p$.*

This means that there exists a function $h$ defined in $[0,1]$ such that

$U$ ("a category's journal with impact factor in the range $[l, l + \Delta l)$ occurs") $= h(p)$

(A.1)

This is a natural property because we assume that subject categories can be characterized by impact-factor histograms (i.e., discrete probabilities) as shown in Figs. 1, 2, and 3.

Next, a second postulate is formulated to obtain a reasonable estimate of how unexpected a subject category was from some impact factor histogram by means of the mathematical expectation of how unexpected its single events were from this same histogram.

Let $p(l=R)$ and $p(l=I)$ be the probability of occurrence of a category's journal with impact factor in the interval $[l, l+\Delta l)$ for a reference category $R$ and the input category $I$, respectively. Suppose that every possible observation from $p(l=R)$ is also a possible observation from $p(l=I)$.

As stated in Axiom 1, if the single events of the reference subject category $R$ are characterized by an "estimated" distribution $Q = \{p(l_i/I) | i = 0, 1, \cdots, n\}$, then the function $h(p(l_i/I))$, with $i = 0, 1, \cdots, n$, returns a measure of how unexpected each single event "a category's journal with impact factor in the interval $[l, l+\Delta l)$ occurs" was from $Q$. Recall that $l_{i+1} = l_i + \Delta l$. Thus, assuming that $P = \{p(l_i/R) | i = 0, 1, \cdots, n\}$ is the "true" probability distribution of the reference subject category $R$, we have that:

**Axiom 2.** *The mathematical expectation of the discrete random variable* $h(Q)$, *which can assume the values*

$$h(p(l_0/I)), h(p(l_1/I)), \cdots, h(p(l_n/I))$$

*with respective probabilities*

$$p(l_0/R), p(l_1/R), \cdots, p(l_n/R)$$

*is an estimate* $U_P(Q)$ *of how unexpected the reference subject category* $R$ *was from* $Q = \{p(l/I)\}$, *i.e.,*

$$U_P(Q) = E_P[h(Q)] = \sum_l p(l/R) h(p(l/I))$$

(A.2)

with $E_P$ denoting the mathematical expectation in $P$.

The following postulate relates the estimate of how unexpected the reference subject category was from an "estimated" distribution and the estimate from the "true" distribution.

**Axiom 3.** *The reference subject category $R$ with "true" probability distribution $P$ is more unexpected from an "estimated" distribution $Q$ than from the \true" distribution $P$.*

The following inequality expresses how the reference category is more unexpected when it is characterized by $Q$ than when is characterized by $P$:

$$U_P(Q) \geq U_P(P)$$

(A.3)

with $U_P(Q)$ and $U_P(P)$ being estimates of how unexpected the reference subject category was from the "estimated" distribution $Q$ and from the "true" distribution $P$, respectively.

That is, here the true distribution $Q$ of the input subject category $I$ can be interpreted as an estimated distribution of the reference category $R$ (with "true" distribution $P$). Thus, we can define a measure of information gain of the reference subject category from the input category by the difference between the estimates of how unexpected the reference subject category was from $Q$ and from $P$.

**Definition: A measure of information gain between subject categories.** Given the reference category $R$ with "true" probability distribution $P = \{p(l/R)\}$, a measure of the information gain of the reference subject category $R$ from the input category $I$ with "true" distribution $Q = \{p(l/I)\}$, is:

$$\varepsilon(P,Q) = U_P(Q) - U_P(P)$$

(A.4)

with $U_P(Q)$ and $U_P(P)$ being estimates of how unexpected the reference subject category was from $Q$ and $P$, respectively. $U_P(Q)$ and $U_P(P)$ are defined as given in Axiom 2, and such that satisfy the inequality (3) in Axiom 3.

The following result serves to determine the form of the measure $\varepsilon(P,Q)$ given in Equation (A.4). It demonstrates that a measure of relative information between two discrete probability distributions, such that satisfies Axioms 1-3, has the form of the Kullback-Leibler information function (Kullback, 1978) as given in Equation (**??**) up to a nonnegative multiplicative constant:

**Theorem.** Let $\varepsilon(P,Q)$ be a measure of information gain for the discrimination between two subject categories as given in equation (A.4) with $P = \{p(l/R)\}$ and $Q = \{p(l/I)\}$. Then, the measure of relative information $E$ is equal to the Kullback-Leibler's information function (Kullback, 1978) between $P$ and $Q$ up to a nonnegative multiplicative constant, i.e.,

$$\varepsilon(P,Q) = aE_P\left[\log \frac{P}{Q}\right]$$

(A.5)

with $a \geq 0$ and $E_P$ denoting the mathematical expectation.

*Proof.* It follows from Theorem 1 in (Garcia et al., 2001).

In conclusion, any measure $\varepsilon(P,Q)$ of how unexpected a subject category was, that satisfies Axioms 1, 2, and 3, has to be of the form of the Kullback-Leibler information function up to a nonnegative multiplicative constant. Hence, the Kullback-Leibler information function is a measure of the information gain between two subject categories, with a minimal number of properties which are natural and thus desirable. It follows that the minimum value of this information gain between two subject categories leads to the most similar ones.